\documentclass[%
 reprint,
 amsmath,amssymb,
 aps,
prb,
]{revtex4-2}

\usepackage{graphicx}
\usepackage{dcolumn}
\usepackage{bm}
\usepackage{hyperref}

\usepackage{appendix}
\usepackage{float}
\usepackage[utf8]{inputenc}

\begin{document}

\preprint{APS/123-QED}

\title{Current-induced Dynamics of Bloch Domain-wall Bimerons}

\author{Jiwen Chen}
\affiliation{School of Science and Engineering, The Chinese University of Hong Kong, Shenzhen, 518172, China}

\author{Laichuan Shen}
\affiliation{School of Science and Engineering, The Chinese University of Hong Kong, Shenzhen, 518172, China}

\author{Oleg A. Tretiakov}%
\email{o.tretiakov@unsw.edu.au}
\affiliation{School of Physics, The University of New South Wales, Sydney 2052, Australia}

\author{Xiaoguang Li}
\email{lixiaoguang@sztu.edu.cn}
\affiliation{College of Engineering Physics, Shenzhen Technology University, Shenzhen, 518118, China}

\author{Yan Zhou}
\email{zhouyan@cuhk.edu.cn}
\affiliation{School of Science and Engineering, The Chinese University of Hong Kong, Shenzhen, 518172, China}



\begin{abstract}
Domain-wall bimerons are composite topological structures formed by embedding bimerons within domain walls in ferromagnets with in-plane anisotropy. These hybrid textures have recently attracted significant attention due to their promise for racetrack memory applications. In this work, we systematically investigate the current-driven dynamics of single domain-wall bimerons and bimeron chains under spin-transfer torque (STT) and spin-orbit torque (SOT). We show that when the spin current is injected or polarized perpendicular to the domain wall, the bimeron Hall effect facilitates efficient motion along the wall. In contrast, spin currents injected or polarized parallel to the wall suppress transverse motion, leading to a significant reduction in the bimeron Hall angle. This anisotropic response is observed for both STT and SOT driving mechanisms. Furthermore, we find that increasing the number of bimerons within a domain wall diminishes their collective mobility. These results provide key insights into domain-wall bimeron dynamics and offer guidance for their integration into bimeron-based spintronic devices.
\end{abstract}

\maketitle


\section{Introduction}
Magnetic skyrmions in perpendicular magnetic anisotropy (PMA) systems and their counterparts, bimerons, in in-plane anisotropy systems have attracted significant attention due to their topological protection, nanoscale dimensions, and high efficiency in current-driven manipulation \cite{mühlbauer2009skyrmion,nagaosa2013topological,gobel2021beyond,yu2012skyrmion,gobel2019magnetic,shen2022nonreciprocal}. These topological spin textures are considered promising candidates for next-generation spintronic devices. However, both skyrmions and bimerons exhibit a transverse deflection known as the skyrmion Hall effect~\cite{nagaosa2013topological, Litzius2017}, which leads to their deviation from the driving direction and potential annihilation at device edges—posing a major challenge for racetrack memory applications. To overcome this limitation, various strategies have been proposed, including the use of topologically compensated spin textures, such as antiferromagnetic skyrmions~\cite{Barker2016, zhang2016antiferromagnetic}, synthetic antiferromagnetic skyrmions~\cite{zhang2016magnetic}, and skyrmioniums~\cite{gobel2019electrical}; increasing the edge energy barrier of nanotracks~\cite{lai2017improved}; and utilizing strip domain wall as a buffer to guide the motion \cite{xing2020enhanced}. 

Recently, composite objects with topological defects embedded in a domain wall have been proposed in various systems \cite{jennings2013dynamics,gudnason2014domain,cheng2019magnetic,li2021field,nagase2021observation,li2021magnetic,lepadatu2020emergence,ross2023domain,amari2024domain,guang2024confined,han2024tunable,xiao2025domain,chen2025magnetic,nie2025current, Amin2023} and are experimentally observed \cite{li2021field,nagase2021observation,lepadatu2020emergence,li2021magnetic, Amin2023}. These structures are referred to as domain-wall skyrmions in PMA systems \cite{jennings2013dynamics,gudnason2014domain,cheng2019magnetic,ross2023domain,amari2024domain,han2024tunable,nie2025current}, and as domain-wall bimerons in systems with in-plane easy axis \cite{nagase2021observation,amari2024domain,chen2025magnetic}. Although domain-wall skyrmions and bimerons retain their topological protection, their motion is effectively confined to the domain wall, which suppresses the skyrmion Hall effect~\cite{chen2025magnetic,nie2025current}, offering a robust solution for future spintronic device architectures.

In this work, we present a comprehensive investigation of dynamics of bimerons embedded in a Bloch-type domain wall in a cubic $\beta$-Mn-type chiral magnet, specifically Co-Zn-Mn thin films. This kind of Bloch domain-wall bimeron and its chain-like texture are observed in \cite{nagase2021observation}. Using micromagnetic simulations based on the Landau-Lifshitz-Gilbert (LLG) equation, in conjunction with Thiele’s collective coordinate approach, we compare the effects of STT and SOT on the dynamics of the domain-wall bimeron. Our results demonstrate that the domain wall not only effectively suppresses the bimeron Hall effect but also allows the Magnus force to serve as the dominant driving mechanism. Moreover, we show that increasing the number of bimerons embedded within the domain wall leads to a reduction in mobility, whereas the spacing between individual bimerons has a negligible impact on their dynamics.

\section{Model and Statics}
In this work, we adopt the material parameters of a cubic $\beta$-Mn-type chiral magnet Co-Zn-Mn thin film \cite{nagase2021observation}, with the free energy density expressed as:
\begin{equation}
\begin{aligned}
    \varepsilon =& A(\nabla\boldsymbol{m})^2 + D\boldsymbol{m}\cdot(\nabla\times\boldsymbol{m}) - \frac{1}{2}M_\text{s}\boldsymbol{m}\cdot\boldsymbol{B}_{\text{demag}} \\
    &+K_c\left[(m_x^2m_y^2 + m_y^2m_z^2 + m_z^2m_x^2\right],
\end{aligned}
\end{equation}
where $\boldsymbol{m}$ is the unit magnetization. The saturation magnetization is $M_\text{s} = 180$ kA/m, the exchange stiffness constant is $A = 3.83$ pJ/m, the bulk Dzyaloshinskii–Moriya interaction (DMI) strength is $D = 0.332$ mJ/$\text{m}^2$, and the cubic magnetic anisotropy constant is $K_c = D^2/4A \approx 7.19\times 10^3$ J/m$^3$. The dipolar field $\boldsymbol{B}_{\text{demag}}$ is included to stabilize the bimeron. Micromagnetic simulations are performed using the open-source software Mumax3 \cite{vansteenkiste2014design}. The magnetization is discretized into a grid mesh of size $512\times 1024\times 1$, with each cell having dimensions of $2\text{ nm}\times 2\text{ nm}\times 2\text{ nm}$. In the $x$-direction, a Neumann boundary condition is applied. In the $y$-direction, periodic boundary conditions are imposed, effectively simulating 21 replicas of the system in this direction. This accounts for the dipolar field generated by these 21 replicated simulation areas \cite{vansteenkiste2014design}. We note that the simulation window is large enough to avoid the effect of the boundary. The simulated temperature is set to zero; however, the simulation results, which account for thermal effects, indicate a thermal stability of up to 100 K. This is discussed in the Appendix~\ref{Append:A} and shown in Fig.~\ref{Fig9}.

\begin{figure}[h]
\includegraphics[width=0.5\textwidth]{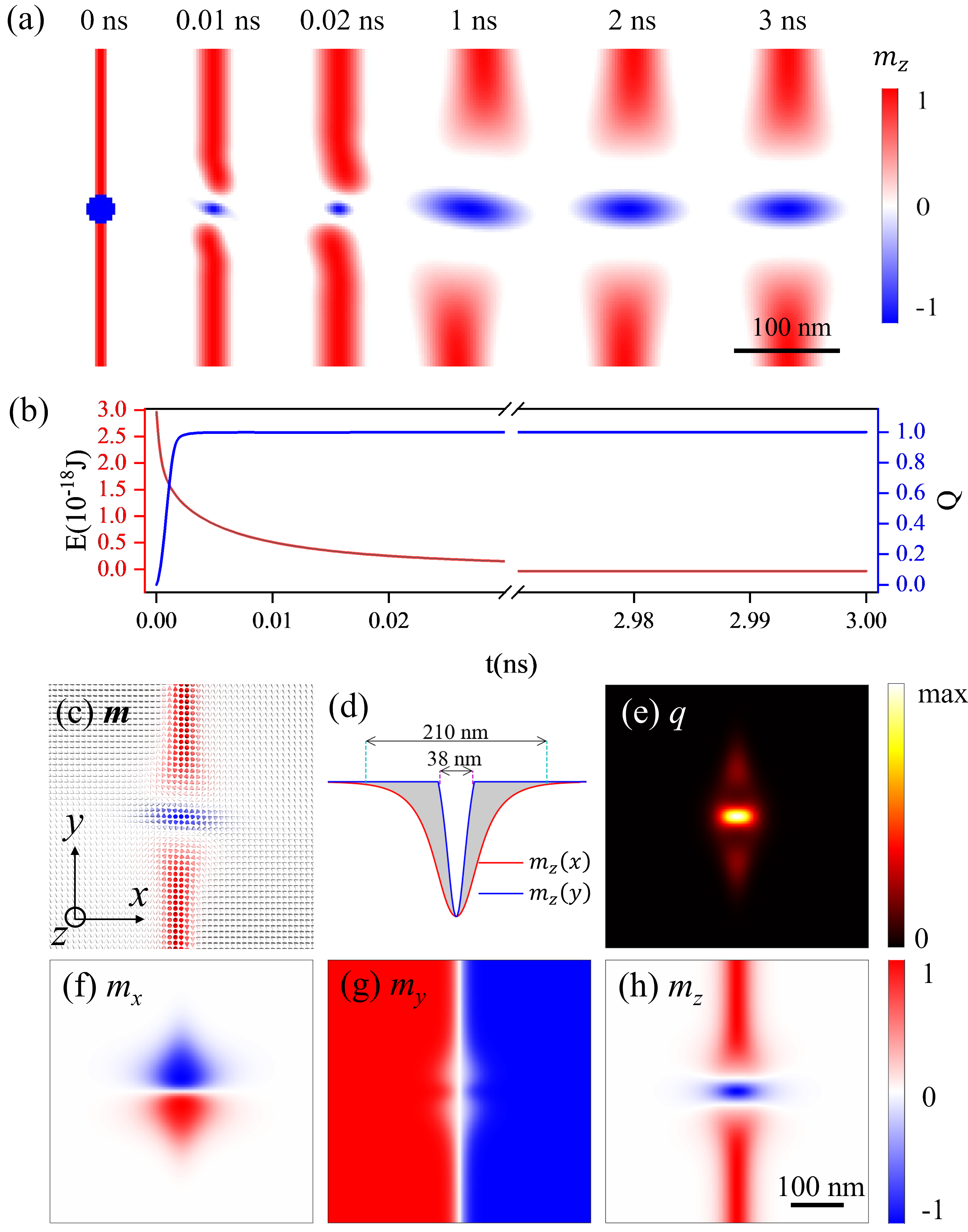}
\caption{\label{Fig1} Panel (a) illustrates the numerical formation of domain-wall bimeron with damping constant $\alpha=0.5$. The initial configuration is a Bloch domain wall with centeral round area being reversed. During the relaxation, the change of topological charge $Q$ and free energy $E$ are presented in panel (b). Panel (c) presents the magnetization vector for domain-wall bimeron and its components are shown in panels (f)-(h). Panel (d) depicts the $m_z$ component of the embedded bimeron's spatial distribution along the $x$- and $y$-directions, defining the bimeron's widths as $r_x = 210$ nm in the $x$-direction and $r_y = 38$ nm in the $y$-direction when $m_z$ transitions from $-1$ to $-0.02$. Panel (e) shows the topological charge density $q$, integrated over the entire plane to yield a total topological charge of +1.}
\end{figure}

Under a zero external field, a bimeron embedded in a Bloch domain wall, referred to as DWBM \cite{nagase2021observation}, can be stabilized, as depicted in Fig.~\ref{Fig1}(a). A Bloch domain wall with a reversed central magnetization region serves as the initial configuration. After 10 fs of relaxation, a bimeron emerges within the domain wall with a topological charge of $Q=+1$ as presented in Fig.~\ref{Fig1}(b). Following 2 ns of relaxation under a damping constant $\alpha=0.5$, the structure becomes fully stabilized, during which the domain wall tends to straighten to minimize the free energy. Constrained by the domain wall, the bimeron adopts an elliptical shape, with a small portion of the topological charge distributed at the end of the domain wall. Under a finite external magnetic field, both the bimeron shape and the charge distribution gradually become more circular and compact, as shown in Fig.~\ref{Fig8}. In the following, we consider the system in the absence of an external magnetic field. Figure \ref{Fig1}(c) shows the magnetization profile, where on the left-hand side it points in the $+y$ direction, while on the right-hand side it points in the $-y$ direction. In the center, the Bloch domain wall (DW) has its magnetization oriented along the $+z$ direction. For the bimeron (BM) region, the $z$-component of the magnetization is negative ($m_z<0$), with the boundary of the BM defined by $m_z \approx 0 $. This results in a size of BM with $r_x=210$ nm in the $x$-direction and $r_y = 38$ nm in the $y$-direction, as indicated in Fig.~\ref{Fig1}(d). The topological charge density, defined as $q(x,y) = \frac{1}{4\pi} \boldsymbol{m}\cdot (\partial_x \boldsymbol{m} \times \partial_y \boldsymbol{m})$ and shown in Fig.~\ref{Fig1}(e), confirms that the BM region possesses a non-zero topological charge. This charge density integrates to $Q=+1$, indicating that the magnetization wraps a unit sphere exactly once \cite{nagaosa2013topological}. The magnetic structure arises from the competition among exchange interaction, DM interaction, magnetic anisotropy, and dipolar interaction. The effects of non-zero external fields and varying strengths of cubic anisotropy on domain-wall bimeron stabilization are explored in the Appendix (see Fig.~\ref{Fig7}). This analysis demonstrates a broad range of parameters that support the stabilization of the DWBM. 

\section{Dynamics driven by spin-transfer torque}
\begin{figure}[h]
\includegraphics[width=0.5\textwidth]{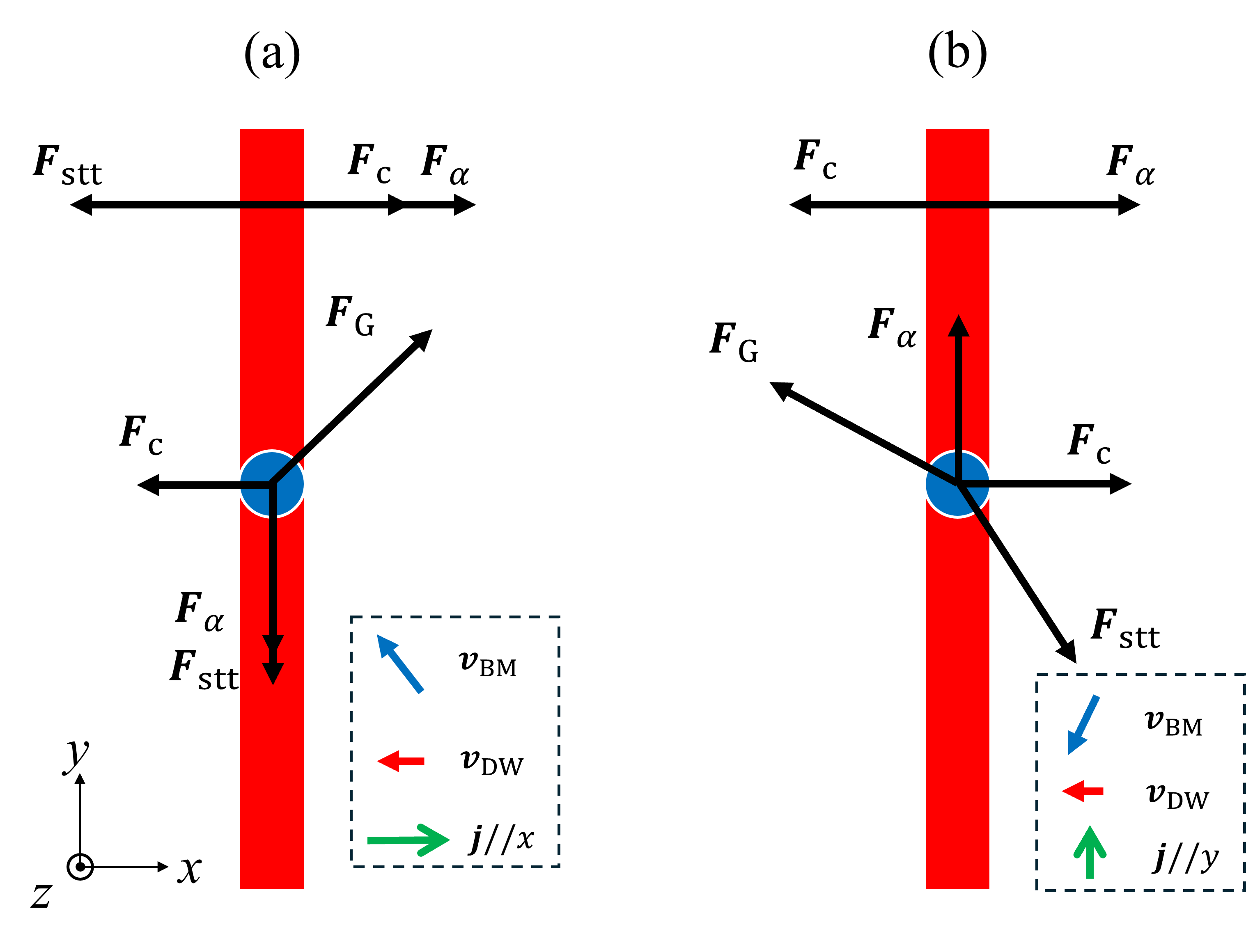}
\caption{\label{Fig2}Force diagrams for current injected in $x$-direction (a) and $y$-direction (b). The red rectangular area represents the domain wall and the blue round area represents the bimeron. $\boldsymbol{F}_\text{G}$ is Magnus force, $\boldsymbol{F}_\alpha$ is damping force, $\boldsymbol{F}_\text{stt}$ is STT driving force and $\boldsymbol{F}_\text{c}$ is constraining force. The parameters used are $\alpha=0.2$ and $j=1\times10^9$ A/m$^2$.}
\end{figure}

We first investigate the dynamics of the domain-wall bimeron by considering spin transfer torque $\tau_\text{STT}$. The STT is described as \cite{zhang2004roles,iwasaki2013current,sampaio2013nucleation}:
\begin{equation}
    \tau_\text{STT}=-(\boldsymbol{v}_s\cdot\nabla)\boldsymbol{m}+\beta\boldsymbol{m}\times(\boldsymbol{v}_s\cdot\nabla)\boldsymbol{m},
\end{equation}
where $\boldsymbol{v}_s=-\gamma\hbar P\boldsymbol{j}/2\mu_0eM_\text{s}(1+\beta^2)$ with $P$ being the spin polarization and $\beta$ being the non-adiabatic STT coefficient. Here $\gamma>0$ is the gyromagnetic ratio, $j$ is the injected current density, $\mu_0$ is the vacuum permeability, and $e$ is the electron's charge. In the simulations, we use spin polarization $P=0.1$ and non-adiabatic STT coefficient $\beta=0.3$. Within Thiele's collective coordinate approach \cite{thiele1973steady, Tretiakov2008, Clarke2008}, we treat the domain wall and the domain-wall bimeron as rigid objects by considering the constraining force between the domain wall and bimeron, which is described in more detail in the Appendix \ref{appendix:thiele}. The motion of domain-wall bimeron under STT driving is determined by \cite{iwasaki2013current,guang2024confined}:
\begin{equation}
\label{eq:STT_thiele}
\boldsymbol{G}\times(\boldsymbol{v}-\boldsymbol{v}_s)-\mathcal{D}\left(\alpha\boldsymbol{v}-\beta\boldsymbol{v}_s\right)=0.
\end{equation}
Here, $\boldsymbol{v}$ is the velocity of the domain-wall bimeron, and the domain wall moves laterally with a velocity $v_x$. The gyromagnetic vector is given by $\boldsymbol{G}=[0,0,-4\pi Q]$, and the damping tensor is defined as $\mathcal{D} = 4\pi D_{\mu\nu}$ with $D_{\mu\nu}=1/4\pi\iint(\partial_\mu\boldsymbol{m}\cdot\partial_\nu\boldsymbol{m})~\text{d}x\text{d}y$. We neglect boundary effects due to the considerable distance from the sample edges. Owing to the configuration's symmetry, the damping tensor components can be numerically evaluated as $D_{xy}=D_{yx}=0$, $D_{xx}=23.88$ and $D_{yy}=1.32$. We also note that $\boldsymbol{v}_s\propto\boldsymbol{j}$ and $v_s=-0.032$ m/s at a current density $j=1\times10^9$ A/m$^2$. For the domain-wall bimeron motion, Thiele's approach reveals that dynamics are governed by the balance of four forces: the Magnus force $\boldsymbol{F}_\text{G} = \boldsymbol{G}\times\boldsymbol{v}$, the damping force $\boldsymbol{F}_\alpha = -\alpha\mathcal{D}\boldsymbol{v}$, the STT driving force $\boldsymbol{F}_\text{stt} = -\boldsymbol{G}\times\boldsymbol{v}_s+\beta\mathcal{D}\boldsymbol{v}_s$, and the constraining force $\boldsymbol{F}_\text{c}$ due to the interaction between the domain wall and the bimeron. We show the force diagram for the current injected along the $x$- and $y$-directions in Fig.~\ref{Fig2} for $\alpha=0.2$ and $j=1\times10^9$ A/m$^2$.

\begin{figure}[h]
\includegraphics[width=0.5\textwidth]{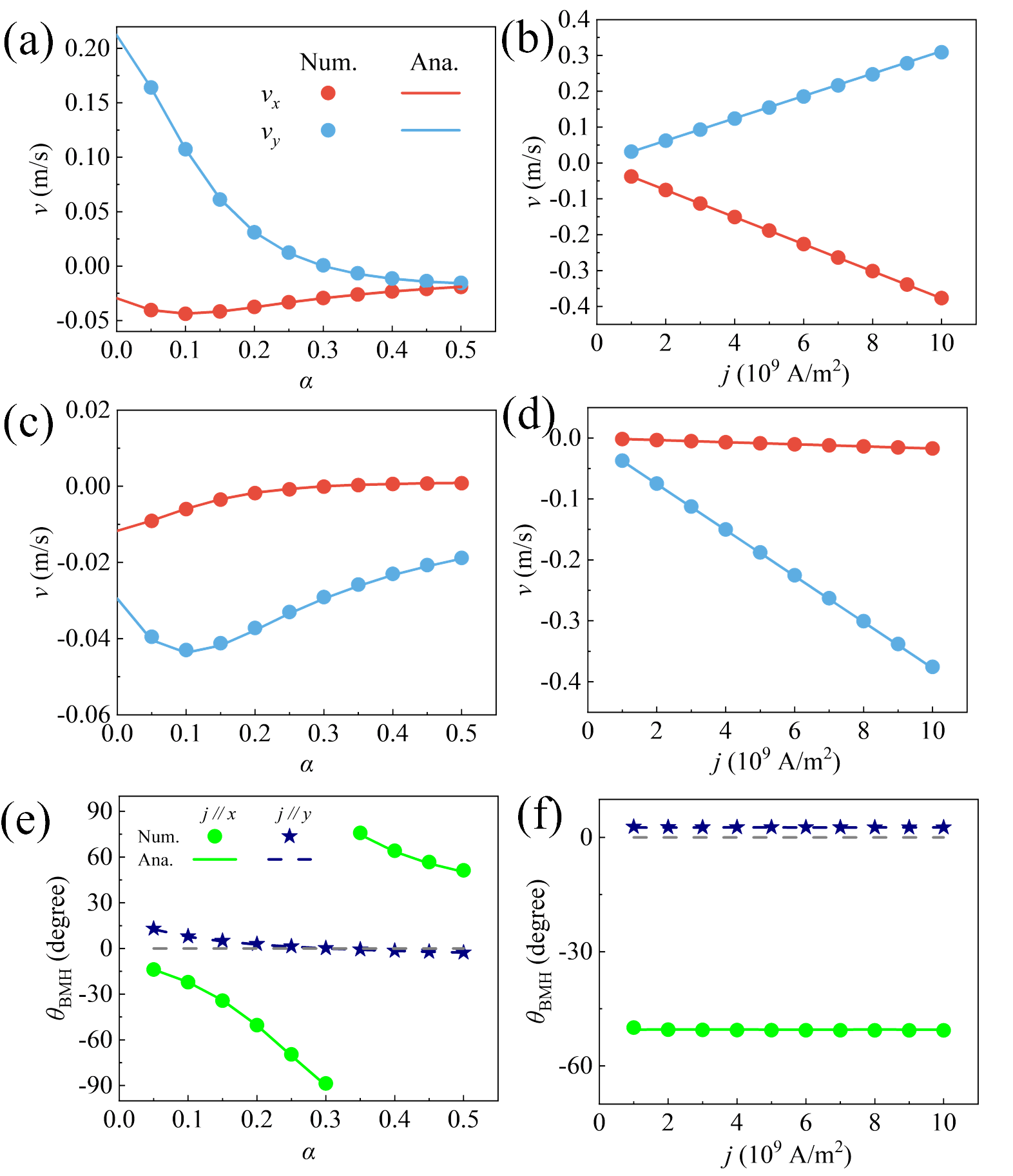}
\caption{\label{Fig3} 
Dynamics of domain-wall bimeron driven by spin-transfer torque with current injected in $x$- and $y$-directions. As for current in $x$-direction, panels (a) and (b) show the velocities as functions of damping constant $\alpha$ and current density $j$ respectively, while panels (c) and (d) show similar results for current injected in $y$-direction. Panels (e) and (f) show the bimeron Hall angle as functions of $\alpha$ and $j$, respectively. The current density is set at $j = 1 \times 10^9$ A/m$^2$ for panels (a), (c) and (e). In panels (b), (d), and (f), the damping constant is fixed at $\alpha = 0.2$. Markers denote numerical results, while solid lines represent the corresponding analytical predictions.
}
\end{figure}

When current is injected in $x$-direction ($\boldsymbol{v}_s = [v_s,0]$), as illustrated in Fig.~\ref{Fig2}(a), the STT drives the domain wall in $-x$-direction and the domain wall bimeron in $-y$-direction. Initially, the domain wall bimeron moves in $-y$-direction and shares the same collective lateral motion with the domain wall due to the constraining force $\boldsymbol{F}_c$. However, as the lateral motion increases, the $y$-direction component of the Magnus force compensates with $\boldsymbol{F}_\text{stt}$ and the Magnus force starts to dominate the dynamics of the domain wall bimeron and results in the motion in $+y$-direction. The Magnus force can significantly enhance the mobility, which is also discussed in \cite{chen2025magnetic,nie2025current}. The velocities of the domain-wall bimeron can be derived from Eq.~(\ref{eq:STT_thiele}) as:
\begin{gather}
    \label{eq:vx_jx}v_x^{\text{STT, } j//x} = \frac{Q^2+\alpha\beta  D_{xx}D_{yy}}{Q^2+\alpha^2D_{xx}D_{yy}}v_{s},  \\
    \label{eq:vy_jx}v_y^{\text{STT, } j//x} = \frac{(\alpha-\beta)QD_{xx}}{Q^2+\alpha^2D_{xx}D_{yy}}v_{s}.
\end{gather}
Here, $j//x$ denotes a current applied along the $x$-direction. These analytical results reveal that the velocity of the domain-wall bimeron exhibits a nonlinear dependence on the damping constant $\alpha$, while scaling linearly with the current density $j$. Notably, when the damping equals the non-adiabatic STT parameter ($\alpha=\beta$), the transverse motion vanishes, i.e., $v_y^{\text{STT, } j//x} = 0$. These analytical predictions are well supported by the numerical simulations presented in Fig.~\ref{Fig3}(a) and (b). The numerical velocities are determined through the topological center 
\begin{equation}
r_i = \frac{1}{Q}\int i \cdot q(x,y)~\mathrm{d}x\mathrm{d}y, \quad  i = x,y.
\end{equation}

When current is injected in the $y$-direction ($\boldsymbol{v}_s = [0,v_s]$), the Magnus and damping forces act together to reduce the mobility of the domain-wall bimeron, as illustrated in Fig.~\ref{Fig2}(b). This response contrasts sharply with the dynamics under 
$x$-direction current. For current applied along the $y$-axis, the domain-wall bimeron’s velocity components are analytically given by:
\begin{gather}
    \label{eq:vx_jy}v_x^{\text{STT, } j//y} = -\frac{D_{yy}}{D_{xx}}v_y^{\text{STT, } j//x},\\ 
    \label{eq:vy_jy}
    v_y^{\text{STT, } j//y} = v_x^{\text{STT, } j//x}.
\end{gather}
The above results demonstrate an antisymmetric relationship between the velocities induced by currents injected along the $x$- and $y$-directions, which originates from the intrinsic antisymmetric structure of the domain-wall bimeron. These analytical predictions are corroborated by numerical simulations, as shown in Figs.~\ref{Fig3}(c) and (d). Compared with the results for current injected in $y$-direction shown in Fig.~\ref{Fig3}(c), a pronounced enhancement in $y$-direction's mobility is observed at low damping, as shown in Fig.~\ref{Fig3}(a), attributed to the effective utilization of the Magnus force as the primary driving mechanism. Furthermore, Fig.~\ref{Fig3}(d) confirms a linear dependence of the bimeron velocity on the current density.

The bimeron Hall angle quantifies the degree of lateral deflection of the bimeron and is defined as $\theta_\text{BMH} = \arctan (v_x/v_y)$. For currents injected along the $x$- and $y$-directions, the Hall angles exhibit distinct behaviors that reflect the anisotropic response of the domain-wall bimeron to different current orientations They are determined by Eqs.~(\ref{eq:vx_jx})-(\ref{eq:vy_jy}):
\begin{align}
    \label{eq:Hall_angle_px}
    \theta_\text{BMH}^{\text{STT, } j//x} &= \arctan \left(\frac{Q^2+\alpha\beta D_{xx} D_{yy}}{(\alpha-\beta)QD_{xx}}\right),\\ 
    \label{eq:Hall_angle_py} 
    \theta_\text{BMH}^{\text{STT, } j//y} &= \arctan \left(-\frac{(\alpha-\beta)QD_{yy}}{Q^2+\alpha\beta D_{xx} D_{yy}}\right).
\end{align}
Figures \ref{Fig3}(e) and (f) present the domain-wall  bimeron Hall angles as functions of damping constant $\alpha$ and current density $j$ for $x$- and $y$-directions, respectively. Under $x$-direction injected current, the bimeron Hall angle is small when decreases the damping constant nearly to 0 and the domain-wall bimeron moves along the domain wall with a significant velocity. Similarly, the bimeron Hall angle goes to 0, when the damping constant equals to $\beta=0.3$ for the scenario of current injected in $y$-direction. These results suggest that the domain wall constrains the bimeron motion, effectively suppressing the Hall effect and guiding its trajectory. The linear dependence on the current density is further confirmed in Fig.~\ref{Fig3}(f), where the bimeron Hall angles are plotted as a function of $j$.

\section{Dynamics driven by spin-orbit torque}
\begin{figure}[h]
\includegraphics[width=0.5\textwidth]{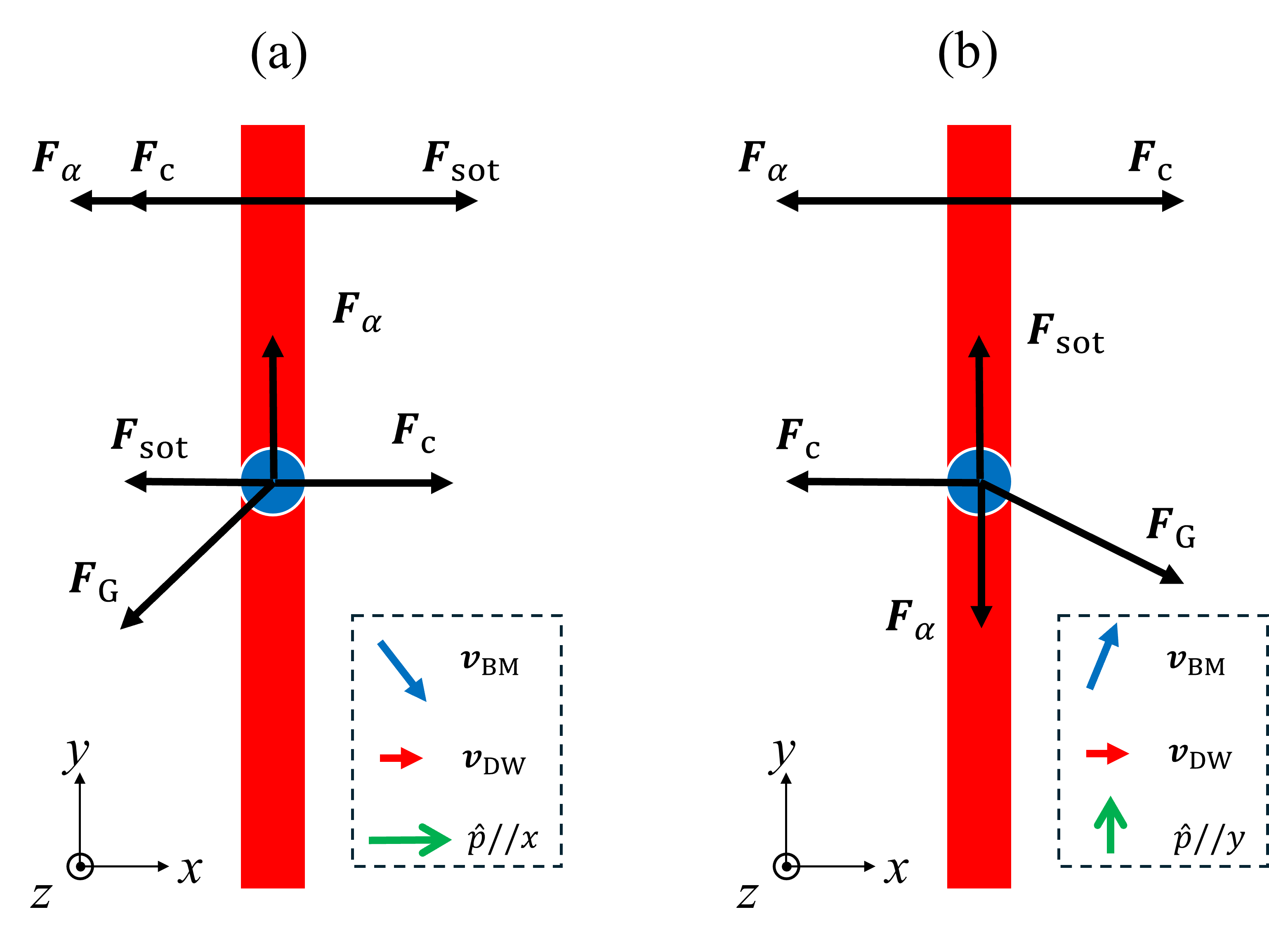}
\caption{\label{Fig4}Force diagrams for spin current polarized in $x$-direction (a) and $y$-direction (b). The red rectangular area represents the domain wall and the blue round area represents the bimeron. $\boldsymbol{F}_\text{G}$ is Magnus force, $\boldsymbol{F}_\alpha$ is damping force, $\boldsymbol{F}_\text{sot}$ is SOT driving force, and $\boldsymbol{F}_\text{c}$ is constraining force. The parameters used are $\alpha=0.2$ and $j=1\times10^9$ A/m$^2$.}
\end{figure}

Here, we focus on the dynamics of domain-wall bimeron driven by the SOT arising from the spin current induced by the spin-Hall effect in an adjacent heavy metal layer \cite{liu2011spin}. Specifically, we consider only the damping-like component of the SOT, expressed as $\tau_\text{SOT}=-\tau_\text{SH}\boldsymbol{m}\times(\boldsymbol{m}\times\hat{p})$, where $\tau_\text{SH}=\gamma\hbar j\theta_\text{SH}/(2\mu_0etM_\text{s})$. Here, $\theta_\text{SH}$ is spin-Hall angle, $t$ is the thickness of the film, and $\hat{p}$ is the unit polarization vector. In the simulation, we use a spin Hall angle $\theta_\text{SH}=0.1$, which determines the conversion rate from electric current to spin polarized current. Following Thiele's collective approach, the motion of the domain-wall bimeron is determined by 
\begin{equation}\label{eq:SOT_thiele}
    \boldsymbol{G}\times\boldsymbol{v} -\alpha\mathcal{D}\boldsymbol{v} -\mathcal{C}\hat{p} = 0.
\end{equation}
The SOT driving tensor is $\mathcal{C}=4\pi C_{\mu\nu}$, where $C_{\mu\nu} = 1/4\pi\iint\tau_\text{SH}(\partial_\mu\boldsymbol{m}\times\boldsymbol{m)}_\nu~\text{d}x\text{d}y$. Providing the symmetry of the magnetization, the tensor components are determined as $C_{xy}=C_{yx}=0, C_{xx}=-7.85$ and $C_{yy}=-0.38$. The SOT driving force $\boldsymbol{F}_\text{sot}$ together with the other forces are illustrated in Fig.~\ref{Fig4}. 

\begin{figure}[h]
\includegraphics[width=0.5\textwidth]{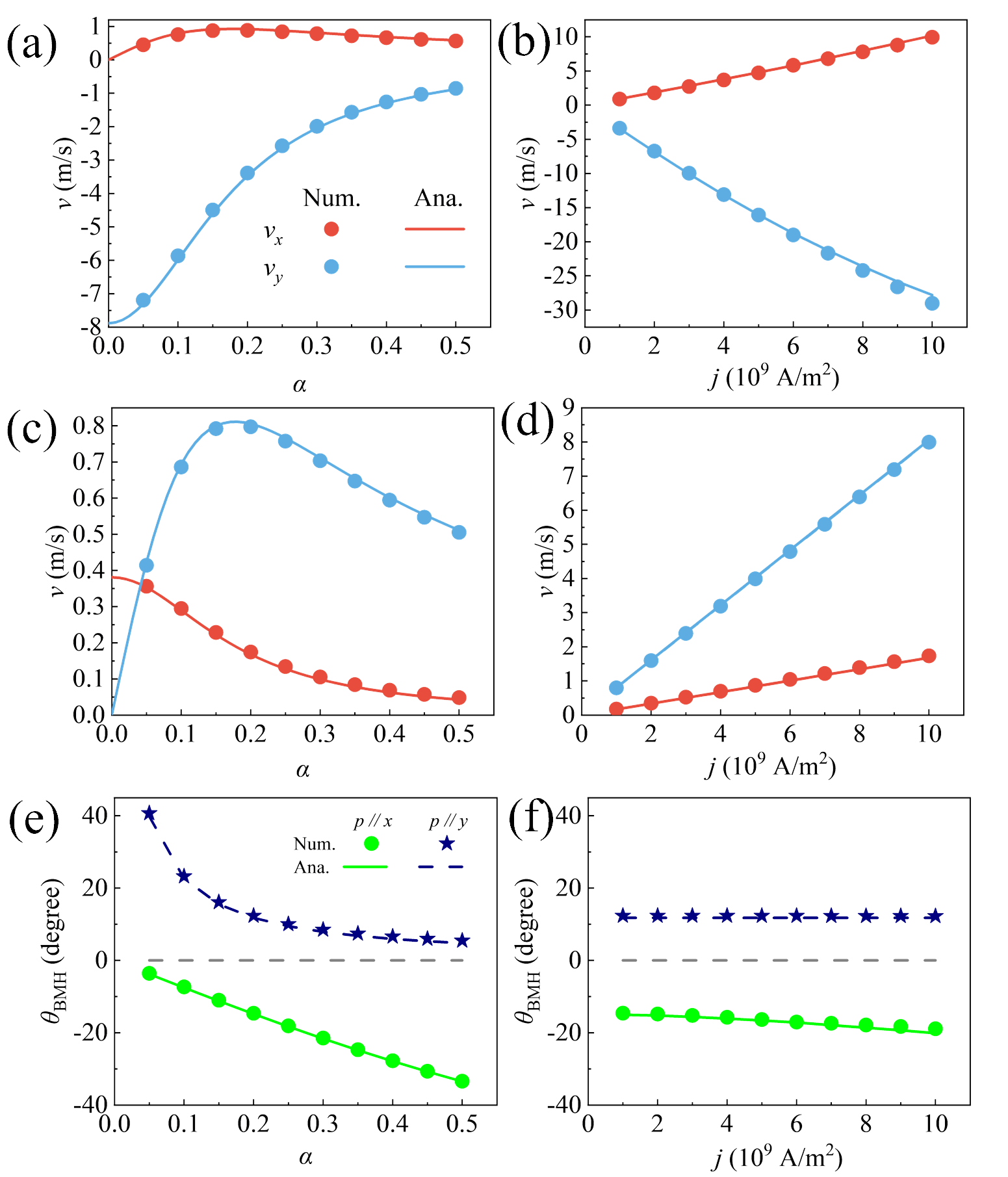}
\caption{\label{Fig5}Dynamics of domain wall bimeron driven by spin orbital torque with spin polarized current injected in $x$- and $y$-directions. As for spin current in $x$-direction, panels (a) and (b) show the velocities as functions of damping constant $\alpha$ and current density $j$ respectively, while panels (c) and (d) show similar results for spin current injected in $y$-direction. The deformation tensors $\mathcal{D}$ and $\mathcal{C}$ are used. Panels (e) and (f) show the bimeron Hall angle as functions of $\alpha$ and $j$, respectively. The current density is set at $j = 1 \times 10^9$ A/m$^2$ for panels (a), (c) and (e). The damping constant is set at $\alpha = 0.2$ for panels (b), (d) and (f). Markers represent numerical results and lines represent analytical results.
}
\end{figure}

As for spin current polarized in the $x$-direction $(\hat{p}=[1,0])$, Fig.~\ref{Fig4}(a) illustrates the forces acting on the domain wall bimeron. Though SOT drives the domain wall bimeron in $-x$-direction, it is counteracted by the constraning force, therefore the domain wall bimeron is embedded into the domain wall. It then moves following the domian wall in $+x$-direction. As a result, the Magnus force intrigues a motion of the domain wall bimeron in $-y$-direction. The velocities of the domain wall bimeron under the driving of spin current polarized $x$-direction are deduced from Eq.~(\ref{eq:SOT_thiele}):
\begin{gather}
    \label{eq:v_px}v_x^{\text{SOT, } \hat{p}//x} = -\frac{\alpha  D_{yy}C_{xx}}{Q^2+\alpha^2D_{xx}D_{yy}},\\  v_y^{\text{SOT, } \hat{p}//x} = \frac{QC_{xx}}{Q^2+\alpha^2D_{xx}D_{yy}}.
\end{gather}
These analytical results are plotted in Fig.~\ref{Fig5}(a) and (b) with numerical results. Compared to the case of STT-driven motion shown in Fig.~\ref{Fig3}(a) and (b), the domain wall bimeron driven by SOT exhibits a pronounced enhancement in mobility—by approximately one order of magnitude—under the same current density and polarization rate ($P = 0.1$ for STT and $ \theta_\text{SH} = 0.1$ for SOT). The driving effect from SOT force is significant. As illustrated in Fig.~\ref{Fig5}(b), the velocity remains linearly dependent on the current density, although a slight deviation appears at higher current densities. This deviation arises from the increasing influence of the Magnus force, which combines with the SOT-induced driving force to compete against the domain wall's confining effect. As a result, local deformation of the domain wall occurs, leading to modifications in both the damping tensor $\mathcal{D}$ and the SOT driving tensor $\mathcal{C}$. In Fig.~\ref{Fig5}, these tensor deformation are incorporated into the analytical calculations, yielding good agreement with the numerical results.  

As for spin current polarized in the $y$-direction $(\hat{p}=[0,1])$, the SOT drives the domain wall bimeron along $y$-direction, as shown in Fig.~\ref{Fig4}(b). The $y$-component of Magnus force impedes the motion of the domain wall bimeron, which is similar to that of STT driving for $y$-direction injected current. From Eq.~(\ref{eq:SOT_thiele}), the domain wall's velocities under the driving of $y$-polarized current become:
\begin{gather}
    v_x^{\text{SOT, } \hat{p}//y} = -\frac{C_{yy}}{C_{xx}}v_y^{\text{SOT, } \hat{p}//x}, \\ v_y^{\text{SOT, } \hat{p}//y} = \frac{D_{xx}C_{yy}}{D_{yy}C_{xx}}v_x^{\text{SOT, } \hat{p}//x}.
\end{gather}
These analytical results are presented in Fig.~\ref{Fig5}(c) and (d) with their corresponding numerical results. As the damping constant increases, the transverse motion of the domain wall bimeron is progressively suppressed, resulting in the motion of BM confined along the domain wall. The bending is negligible, as shown in Fig.~\ref{Fig5}(d), owing to the reduced mobility and consequently weaker Magnus force.

The bimeron Hall angles for both spin current polarized in $x$- and $y$-direction are  
\begin{align}
    \label{eq:Hall_angle_px}
    \theta_\text{BMH}^{\text{SOT, } \hat{p}//x} &= \arctan \left(-\frac{\alpha  D_{yy}}{Q}\right),\\ 
    \label{eq:Hall_angle_py} 
    \theta_\text{BMH}^{\text{SOT, } \hat{p}//y} &= \arctan \left(\frac{Q}{\alpha D_{xx}}\right).
\end{align}
These results are shown in Fig.~\ref{Fig5}(e) and (f). For spin polarized current injected in the $x$-direction, the domain wall bimeron moves rapidly with a small Hall angle, primarily driven by the Magnus force. In contrast, when the current is injected along the $y$-direction, the Magnus force impedes the motion; however, a small bimeron Hall angle can still be achieved at sufficiently high damping. The bending effect is also evident in Fig.~\ref{Fig5}(f), where the bimeron Hall angle slightly increases with current density in the $x$-polarization case, while it remains nearly constant for the $y$-polarization case.



\section{Dynamics of domain-wall bimeron's chain}
\begin{figure*}[h]
\includegraphics[width=0.9\textwidth]{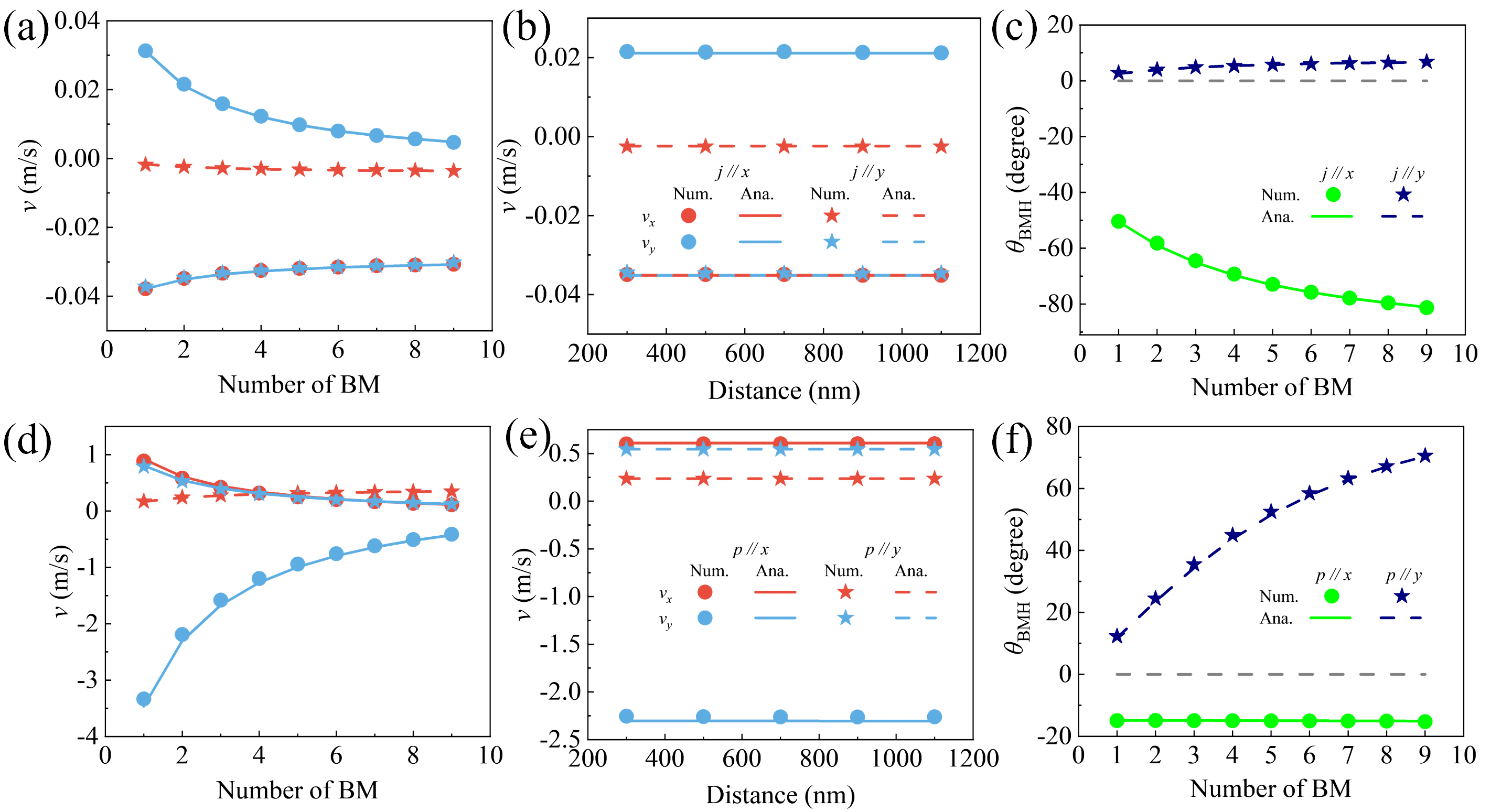}
\caption{\label{Fig6} Dynamics of the domain wall bimerons chain driven by spin transfer torque (a)-(c) and spin orbital torque (d)-(f). Panels (a) and (d) depict the velocities of domain wall bimerons as functions of their number. Panels (b) and (e) illustrate the velocities of only two bimerons embedded in the domain wall as functions of the distance between them.  Panels (c) and (f) shows the bimeron Hall angle $\theta_\text{BMH}$ of the domain wall bimerons chain as functions of their number. The damping constant is set to $\alpha = 0.2$ and the injected current density is $j = 1 \times 10^9$ A/m$^2$.}
\end{figure*}

In this part, we investigate the dynamics of the domain-wall bimeron's chain driven by STT and SOT. The chain consists of an array of equidistant bimerons embedded within a single domain wall and moves as a rigid body without deformation. Its dynamics can be captured by Thiele’s equations [Eqs.~(\ref{eq:STT_thiele}) and (\ref{eq:SOT_thiele})], with the same formalism as in the single bimeron case, but with modified tensor components. The topological charge $Q$ and the tensor elements $D_{yy}$ and $C_{yy}$ are proportional to the number of bimerons, as they depend only on the bimeron region. We validate our analytical predictions through numerical simulations, with results shown in Fig.~\ref{Fig6}(a)–(c) for STT driving and Fig.~\ref{Fig6}(d)–(f) for SOT driving.

Figure~\ref{Fig6}(a) shows the velocities of the domain wall bimeron chain as a function of the number of embedded bimerons driven by STT. The presence of bimerons reduces the overall mobility of the chain for both current injection directions. Moreover, as the number of bimerons increases, the bimeron Hall angle grows accordingly, as shown in Fig.~\ref{Fig6}(c), indicating an enhanced Hall effect induced by the bimerons. When only two bimerons are embedded, their separation distance has negligible influence on dynamics, as demonstrated in Fig.~\ref{Fig6}(b).

Under SOT driving, overall mobility generally decreases with the increase in the number of bimerons, as shown in Fig.~\ref{Fig6}(d). An exception is observed for spin current polarized in $y$-direction, where the transverse velocity increases slightly. The bimeron Hall angle remains nearly constant in this case, which can be attributed to the relation $D_{yy} \propto Q$, leading to a constant value of $\theta_\text{BMH}^\text{SOT, $\hat{p} // x$}$. Similar to the STT case, the distance between bimerons does not affect the dynamics, as shown in Fig.~\ref{Fig6}(e).

\section{Conclusions}
In conclusion, we have comprehensively investigated the statics and dynamics of Bloch domain-wall bimerons by numerically solving the Landau–Lifshitz–Gilbert equation and employing Thiele's analytical approach. The analytical derivation of bimeron dynamics shows good agreement with the miscromagnetic simulations. Our numerical results demonstrate that domain-wall bimeron is robust across a wide range of parameters and can tolerate thermal disturbances up to 100 K. 

We compared the driving mechanisms of spin-transfer torque and spin-orbit torque, revealing that SOT provides significantly higher driving efficiency. Under both torque mechanisms, the Magnus force can be harnessed to drive the bimeron when the current is injected along the \(x\)-direction for STT (or polarized along the \(x\)-direction for SOT). Meanwhile, when the current is injected along the \(y\)-direction for STT (or polarized along the \(y\)-direction for SOT), the domain wall structure effectively suppresses the bimeron's lateral motion, resulting in a small bimeron Hall angle.

Furthermore, we have studied the dynamics of the domain-wall bimerons chain and found that increasing the number of bimerons reduces overall mobility while enhancing the bimeron Hall effect. We also revealed that the spacing between the bimerons has a negligible influence on the dynamics. These findings offer valuable insights into the controlled manipulation of domain-wall bimerons and lay a solid foundation for their potential application in next-generation spintronic devices.

\section*{ACKNOWLEDGMENTS}

O.A.T. acknowledges support from the Australian Research Council (Grant No.~DP240101062), NCMAS grant, and visiting program of ICC-IMR, Tohoku University (Japan). X. L. acknowledges the support from the National Natural Science Foundation of China (Grant No. 12104322), Guangdong Basic and Applied Basic Research Foundation (Grant No. 2025A1515011895), and the Natural Science Foundation of Top Talent of SZTU (Grant No. GDRC202309). Y. Z. acknowledges the support from the Shenzhen Peacock Group Plan (KQTD20180413181702403), the Shenzhen Fundamental Research Fund (Grant No. JCYJ20210324120213037), the Guangdong Basic and Applied Basic Research Foundation (Grant No. 2021B1515120047), the National Natural Science Foundation of China (Grant No. 2374123, 11974298).

\appendix
\section{Stability}
\label{Append:A}
\subsection{Phase Diagram}
\begin{figure*}[h]
\includegraphics[width=1.0\textwidth]{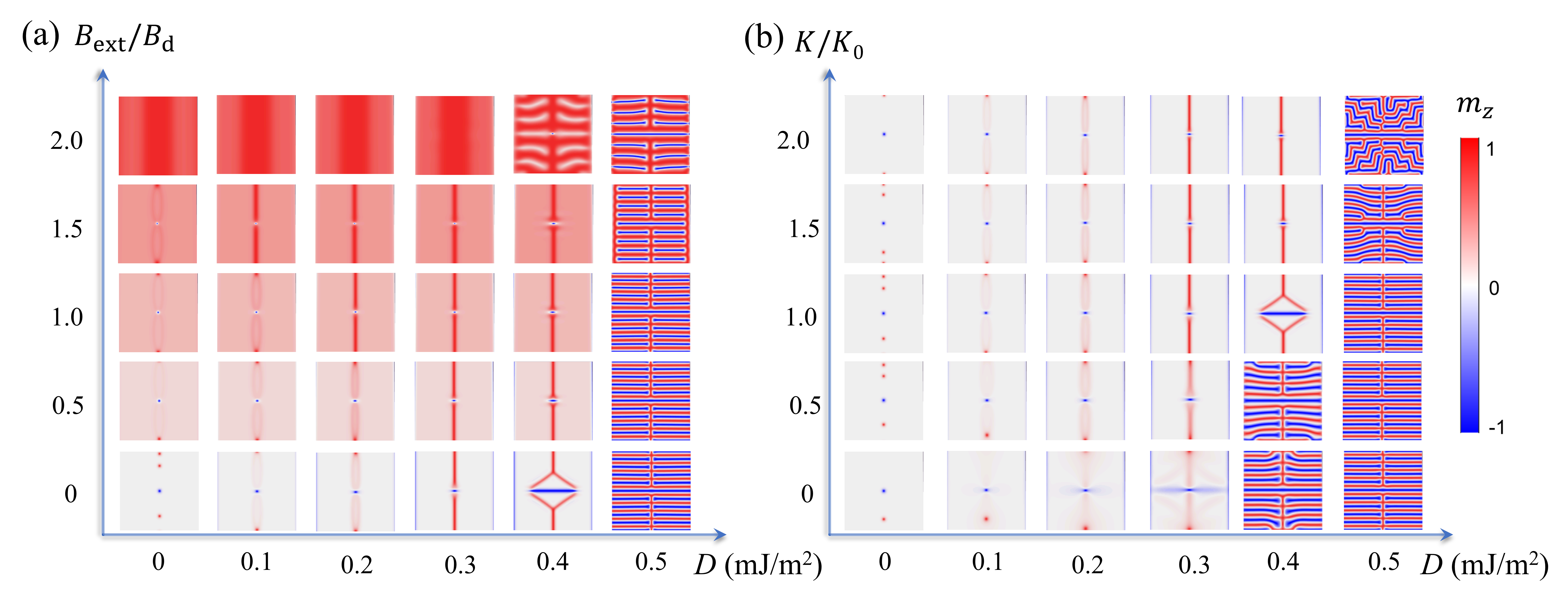}
\caption{\label{Fig7} (a) Phase diagram as function of bulk DMI and external field $B_\text{ext}$ at anistropy constant $K=K_0$; (b) Phase diagram as function of bulk DMI and anisotropy constant $K$ at zero external field.  The perpendicular external magnetic field is $B_{\text{ext}}$ and $B_\text{d} = D^2/2M_\text{s}A\approx 79.94 \text{ mT}$}
\end{figure*}

To explore the conditions for the existence of the domain wall bimeron, we perform numerical simulations by varying the bulk DMI constant $D$ and the external magnetic field $B_\text{ext}$, as shown in Fig.~\ref{Fig7}(a). At zero external field, merons appear as the ground state at small $D$. When $D$ increases at around 0.3 mJ/m$^2$, the domain wall bimeron begin to form. With increasing external field, the shape of the bimeron embedded in the domain wall evolves from elliptical to circular. At large DMI strength, the system eventually transitions into a spiral state. Figure~\ref{Fig7}(b) presents the phase diagram as a function of the DMI constant $D$ and the anisotropy constant $K$ at zero external field. When the DMI and the anisotropy are sufficiently strong (e.g., $K = K_0$), the domain wall bimeron remains stable though a larger DMI favors spiral state. 

\begin{figure}[h]
\includegraphics[width=0.48\textwidth]{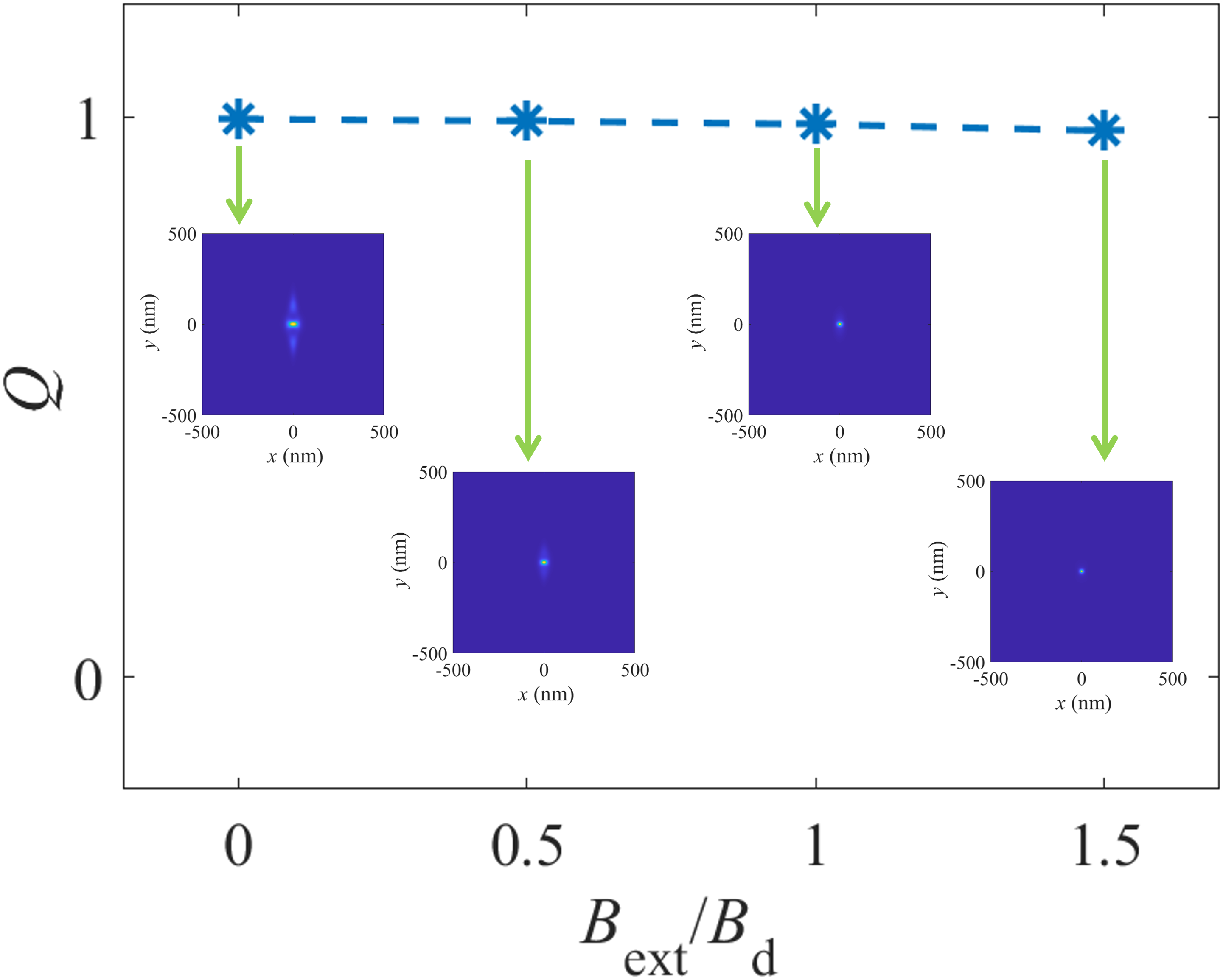}
\caption{\label{Fig8}The topological charge as function of external field $B$ when DMI constant $D=0.3$ mJ/m$^2$. The insets show the topological charge distribution. }
\end{figure}

At a fixed DMI constant of $D = 0.3$ mJ/m$^2$, we vary the external magnetic field from zero to $1.5B_\text{d}$ and present the corresponding topological charge and charge density in Fig.~\ref{Fig8}. The results show that, under the mediation of the external field, the topology of the domain wall bimeron becomes increasingly compact.

\subsection{Effect of Temperature}
\begin{figure}[h]
\includegraphics[width=0.48\textwidth]{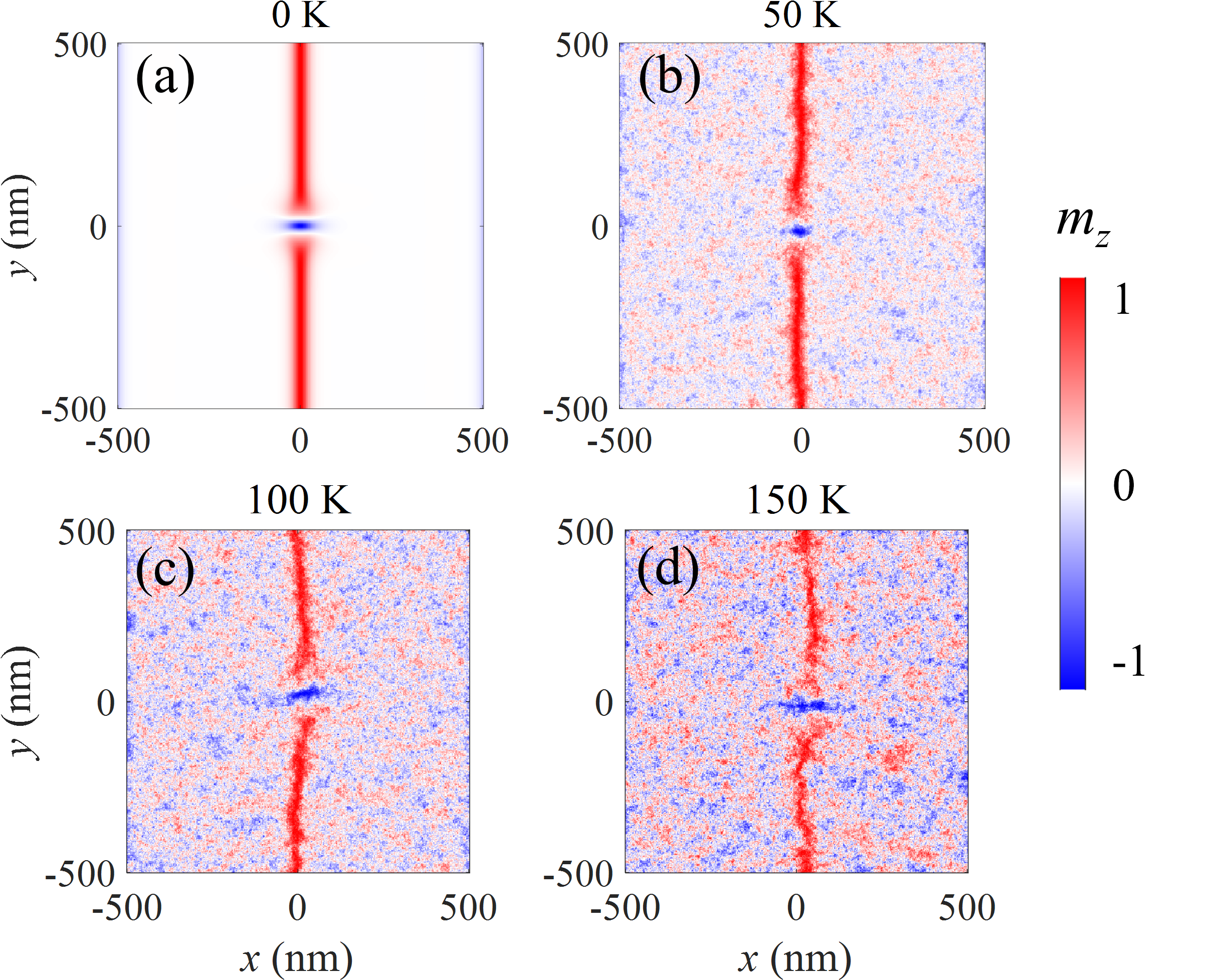}
\caption{\label{Fig9}The thermal stability of the domain wall bimeron with temperature from 0 K to 150 K. These snaps present the z-component magnetization after 30 ns relaxation with damping constant $\alpha = 0.2$. }
\end{figure}

Figure \ref{Fig9} illustrates the thermal stability by implementing dynamics at finite temperature for the same parameters as used in the main text. It is observed that the bimeron remains stable up to 100 K. When the temperature is 150 K, the bimeron deforms significantly.

\section{\label{appendix:thiele}Thiele's Equations}
Under the influence of STT and SOT, the bimeron slides along the domain wall, while the domain wall undergoes lateral motion. To analyze these dynamics, we employ Thiele's equation \cite{thiele1973steady, Tretiakov2008, Clarke2008} with a constraining force $\boldsymbol{F}^\text{c}$ for both the domain wall and the bimeron, separately:
\begin{gather}
    \label{eq:thiele_DW}-\mathcal{D}^\text{DW}\left(\alpha\boldsymbol{v}^\text{DW}-\beta \boldsymbol{v}_s\right)-\mathcal{C}^\text{DW}\hat{p}+\boldsymbol{F}^\text{c}=0, \\
    \label{eq:thiele_BM}\boldsymbol{G}\times\left(\boldsymbol{v}^\text{BM}-\boldsymbol{v}_s\right)-\mathcal{D}^\text{BM}\left(\alpha\boldsymbol{v}^\text{BM}-\beta \boldsymbol{v}_s\right)-\mathcal{C}^\text{BM}\hat{p}-\boldsymbol{F}^\text{c}=0.
\end{gather}
Here $\boldsymbol{v}$ represents the velocity, while ``DW" and ``BM" denote the domain wall and the bimeron region, respectively. The conduction electron velocity is $\boldsymbol{v}_s=-\gamma\hbar P\boldsymbol{j}/2\mu_0eM_\text{s}(1+\beta^2)$. The gyromagnetic vector is given by $\boldsymbol{G}=[0,0,-4\pi Q]$, where $Q$ is the topological charge. The damping tensor is defined as $\mathcal{D} = 4\pi D_{\mu\nu}$ with $D_{\mu\nu}=1/4\pi\iint(\partial_\mu\boldsymbol{m}\cdot\partial_\nu\boldsymbol{m})~\text{d}x\text{d}y$. The SOT driving tensor is $\mathcal{C}=4\pi C_{\mu\nu}$, where $C_{\mu\nu} = 1/4\pi\iint\tau_\text{SH}(\partial_\mu\boldsymbol{m}\times\boldsymbol{m)}_\nu~\text{d}x\text{d}y$. The bimeron shares the same lateral velocity as the domain wall $(v^\text{BM}_x=v^\text{DW}_x)$, while the domain wall has zero velocity in the $y$-direction $(v^\text{DW}_y=0)$. Additionally, the damping tensor for the domain wall has only one non-zero component, $\mathcal{D}^\text{DW}_{xx}$, leading to the relationship $\mathcal{D}^\text{DW}\boldsymbol{v}^\text{DW}=\mathcal{D}^\text{DW}\boldsymbol{v}^\text{BM}$. Using these simplifications, Eqs.~(\ref{eq:thiele_DW}) and (\ref{eq:thiele_BM}) can be reduced to a single equation of motion for the bimeron:
\begin{equation}
\label{eq:thiele}\boldsymbol{G}\times(\boldsymbol{v}-\boldsymbol{v}_s)-\mathcal{D}\left(\alpha\boldsymbol{v}-\beta\boldsymbol{v}_s\right)-\mathcal{C}\hat{p}=0.
\end{equation}
Here $\boldsymbol{v}=\boldsymbol{v}^\text{BM}$, $\mathcal{D}=\mathcal{D}^\text{DW}+\mathcal{D}^\text{BM}$, and $\mathcal{C}=\mathcal{C}^\text{DW}+\mathcal{C}^\text{BM}$. By considering the symmetry of the magnetization, the cross-terms disappear, i.e., $D_{xy}=D_{yx}=C_{xy}=C_{yx}=0$.

\bibliography{Bloch_DWBM}

\end{document}